\documentclass[prb,aps,floats,twocolumn]{revtex4}
\usepackage{epsfig,amssymb,amsmath}

\begin{document}
\title{Introduction to the Bethe Ansatz III} 
\author{Michael Karbach$^*$, Kun Hu$^\dagger$, and Gerhard M{\"u}ller$^\dagger$} 
\affiliation{$^*$Bergische Universit{\"a}t Wuppertal, Fachbereich Physik, 
         D-42097 Wuppertal, Germany \\ 
$^\dagger$Department of Physics, University of Rhode Island, Kingston RI 02881-0817}
\date{\today~--~3.0}
\begin{abstract}
  Having introduced the magnon in part I and the spinon in part II as the
  relevant quasi-particles for the interpretation of the spectrum of low-lying
  excitations in the one-dimensional (1D) s=1/2 Heisenberg ferromagnet and
  antiferromagnet, respectively, we now study the low-lying excitations of the
  Heisenberg antiferromagnet in a magnetic field and interpret these collective
  states as composites of quasi-particles from a different species . We employ
  the Bethe ansatz to calculate matrix elements and show how the results of such
  a calculation can be used to predict lineshapes for neutron scattering
  experiments on quasi-1D antiferromagnetic compounds.  The paper is designed as
  a tutorial for beginning graduate students. It includes 11 problems for
  further study.
\end{abstract}
\pacs{??}
\maketitle
%
\section{Introduction}\label{sec:I}
%
In most areas of condensed-matter research, a few model systems outshine all
others by their prototypical significance, which promises to encapsulate the
essence of a physical phenomenon with little interference of collateral degrees
of freedom. In cooperative magnetism, the Ising model, the Heisenberg model, and
the Hubbard model, for example, are theoretical many-body systems that have kept
generations of researchers fascinated.

The prominence of such models and the abundance of theoretical predictions for
many of their physical properties have at times led to a reversal of the
traditional relationship between theory and experiment. Instead of theorists
proposing and solving models for the purpose of explaining measurements
performed on specific materials, we see magneto-chemists and condensed-matter
experimentalists at work searching for materials which are physical realizations
of the prototypical models.

The focus here is on the one-dimensional (1D) $s=\frac{1}{2}$ Heisenberg
antiferromagnet in a magnetic field. The Hamiltonian for a cyclic chain with $N$
sites reads
\begin{equation}\label{eq:Hh}
H = \sum_{n=1}^N \left[J{\bf S}_n \cdot {\bf S}_{n+1}  - hS_n^z\right].
\end{equation}
This model is amenable to exact analysis via Bethe ansatz and displays dynamical
properties of intriguing complexity. The magnetic field $h$ is a continuous
parameter which has a strong impact on most physical properties and which is
directly controllable in experiments. The high degrees of computational and
experimental control is what makes this system so attractive to researchers.

Physical realizations of Heisenberg antiferromagnetic chains have been known for
many years in the form of 3D crystalline compounds with quasi-1D exchange
coupling between magnetic ions. The desired properties of the best candidate
material include the following: The coupling of the effective electron spins
must be highly isotropic in spin space and overwhelmingly predominant between
nearest-neighbor magnetic ions along one crystallographic axis. The intra-chain
coupling must not be too weak or else it will be hard to study the
low-temperature properties, which are of particular interest. It must not be too
strong either or else it will be hard to reach a magnetic field that makes the
Zeeman energy $hS_n^z$ comparable to the exchange energy $J{\bf S}_n \cdot {\bf
  S}_{n+1}$.

One compound that fits the bill particularly well is {\it copper pyrazine
  dinitrate} [Cu(C$_4$H$_4$N$_2$)(NO$_3$)$_2$], a material that was only
recently synthesized in single crystals of sizes and shapes suitable for
magnetic inelastic neutron scattering.\cite{HSR+99} The most detailed dynamical
experimental results to date were made on KCuF$_3$,\cite{NTC+91} in which the
intra-chain coupling is considerably stronger, which makes it easier to study
low-temperature effects but harder to study magnetic-field effects.

What happens in a magnetic neutron scattering experiment? A beam of
monochromatic neutrons, ideally a plane wave with well defined momentum and
energy, is scattered inelastically off an array of spin chains via magnetic
dipolar interaction between the exchange coupled electron spins and the neutron
spin into a superposition of waves with a range of momenta and energies.  Each
inelastic scattering event causes a transition between two eigenstates of the
spin chain. The difference in energy and wave number of the two eigenstates
involved must be matched by the energy and momentum transfer of the scattered
neutron.

If the measurement is performed at very low temperature, it is reasonable to
assume that the observable scattering events predominantly involve transitions
from the ground state $|G\rangle$ to excitations $|\lambda\rangle$ with energies and wave numbers
within a window preset by the experimental setup. The experiment thus enables us
to have a direct look at the excitation spectrum of the spin chain.  It probes
the spin fluctuations as described by the fluctuation operator
\begin{equation}\label{eq:sqmu}
S_q^\mu = N^{-1/2}\sum_{n=1}^Ne^{iqn}S_n^\mu, \quad \mu=x,y,z
\end{equation}
for wave numbers $q=2\pi l/N, l=1,\ldots,N$. Under idealized circumstances, the
inelastic neutron scattering cross section is proportional to the dynamic spin
structure factor at zero temperature:\cite{BL89,Furr00}
\begin{equation}\label{eq:dssf}
S_{\mu\mu}(q,\omega) = 2\pi\sum_\lambda|\langle G|S_q^\mu|\lambda\rangle|^2
\delta\left(\omega-\omega_\lambda \right).
\end{equation}

Each scattering event with energy transfer $\hbar\omega_\lambda\equiv E_\lambda-E_G$ and momentum transfer
$\hbar q\equiv \hbar(k_\lambda-k_G)$ along the chain induces a transition from $|G\rangle$ to $|\lambda\rangle$ and
contributes a spectral line of intensity $2\pi|\langle G|S_q^\mu|\lambda\rangle|^2$ to $S_{\mu\mu}(q,\omega)$.
For a macroscopic system, the spectral lines are arranged in a variety of
patterns in $(q,\omega)$-space, including branches, continua, and more complicated
structures. The functions $S_{\mu\mu}(q,\omega)$ thus provide a palette of information
about the $T=0$ dynamics of the spin chain.

Unlike in a classical dynamical system, where all motion grinds to a halt at
$T=0$, the quantum wheels keep turning -- a phenomenon known as zero point
motion. For the Heisenberg model (\ref{eq:Hh}) in the ground state $|G\rangle$, the
quantum fluctuations of any observable ${\cal O}$ of interest can be described
by a time-dependent correlation function of the form
\begin{equation}\label{eq:dcf}
\langle {\cal O}(t){\cal O}^\dagger\rangle \equiv 
\langle G|e^{iHt/\hbar}{\cal O}e^{-iHt/\hbar}{\cal O}^\dagger|G\rangle.
\end{equation}
If ${\cal O}$ is conserved ($[{\cal O},H]=0$), then $\langle {\cal O}(t){\cal O}^\dagger\rangle=\langle
G|{\cal O}{\cal O}^\dagger|G\rangle$ is a constant.  Different dynamical variables ${\cal
  O}$ yield different correlation functions for the same ground state $|G\rangle$.
The dynamical variable seen by neutrons during the magnetic scattering is
$S_q^\mu$. Fourier transforming (\ref{eq:dcf}) with ${\cal O}=S_q^\mu$
yields the dynamic spin structure factor (\ref{eq:dssf}) (Problem~1).

Light scattering, electron scattering, photoemission, and nuclear magnetic
resonance are some other experimental techniques used to investigate the
dynamics of (\ref{eq:Hh}). Different measuring techniques view the same
excitation spectrum through lenses of different color, i.e. by transition rates
specific to particular fluctuation operators. Hence each experimental probe
filters out a specific aspect of the zero point motion by viewing a particular
dynamical variable of one and the same system.

The goal set for this column is to teach the reader how to make detailed
predictions based on the Bethe ansatz solution of (\ref{eq:Hh}) for the
inelastic neutron scattering cross section as can be measured on quasi-1D
antiferromagnetic materials.  Many of the computational and analytic tools that
are needed for this purpose were introduced in parts I\cite{KM97} and
II\cite{KHM98} of this series. The calculation of matrix elements from Bethe
wave functions will be introduced here along the way.

%
\section{Quasi-particles at $\lowercase{h}=0$: spinons}
\label{sec:2mse}
%
In part II we used the Bethe ansatz to describe the ground state $|A\rangle$ of
\eqref{eq:Hh} in zero field. We saw that this singlet $(S_T=0)$ state can be
interpreted as the physical vacuum of a particular species of particles with
spin $\frac{1}{2}$: the spinons.\cite{FT81} To distinguish them from the
electrons, protons, and neutrons -- the constituent elementary particles of all
materials -- we use the name quasi-particle for the spinons as is custom.  We
identified a set of low-lying excitated states containing two spinons with spins
up. The spin $S_T^z=S_T=1$ of a (stationary) 2-spinon state is shared by all
electrons of the system in what is called a collective excitation.

\begin{figure}[t!]
  \vspace*{-0.4cm}

\centerline{\epsfig{file=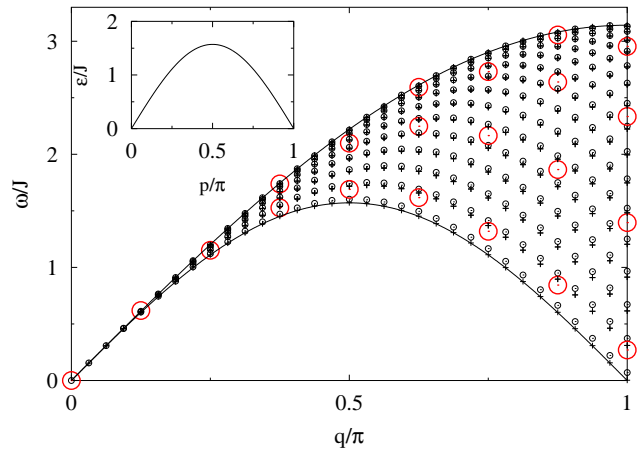,width=6.9cm,angle=-90}}
\caption{Energy versus wave number of all 2-spinon scattering states with
  $S_T^z=S_T=1$ at $q\geq 0$ for $N=64$ ($\circ$) in comparison with the corresponding
  free 2-spinon superpositions (+). The large red circles are 2-spinon data for
  $N=16$. The inset shows the energy-momentum relation \eqref{eq:spdisp} of the
  spinon quasi-particle.}
\label{fig:0}
\end{figure}

Here we revisit these 2-spinon excitations with our eyes focused on the
quasi-particles. The red circles in Fig.~\ref{fig:0} represent the 2-spinon
states for $N=16$ (see also Fig.~4 of part II). We know that for $N\to\infty$ they
close up to form a continuum in $(q,\omega)$-space with boundaries\cite{DP62,Yama69}
\begin{equation}\label{eq:epslu}
\epsilon_L(q) = \frac{\pi}{2}J|\sin q|, \quad \epsilon_U(q) = \pi
J\left|\sin\frac{q}{2}\right|,
\end{equation}
represented by the solid lines in Fig.~\ref{fig:0}.  In every eigenstate of this
set, the spinons can be thought of as two localized perturbations of the spinon
vacuum $|A\rangle$ moving around the chain with momenta $p_1, p_2$ and energies
$\epsilon_{sp}(p_1), \epsilon_{sp}(p_2)$. The spinon energy-momentum relation,
\begin{equation}\label{eq:spdisp}
\epsilon_{sp}(p) = \frac{\pi}{2}J\sin p,\quad 0\leq p\leq\pi,
\end{equation}
is shown in the inset to Fig.~\ref{fig:0}.  Periodically, the two
quasi-particles scatter off each other, hence the name 2-spinon scattering
state.
 
The energies and wave numbers of the 2-spinon eigenstates for $N=64$ obtained
via Bethe ansatz are represented by circles $(\circ)$ in Fig.~\ref{fig:0}, whereas
the corresponding (fictitious) free 2-spinon superpositions with wave numbers
$q=p_1+p_2$ and energy $\omega=\epsilon_{sp}(p_1)+\epsilon_{sp}(p_2)$ are shown as $(+)$ symbols.
The upward displacement of the former relative to the latter is a measure of the
positive spinon interaction energy caused by a predominently repulsive force
acting between the two quasi-particles.

With increasing $N$, the scattering events between the two spinons in a 2-spinon
state become scarcer at a rate inversely proportional to the distance covered by
the quasi-particles between collisions.  Consequently, the spinon interaction
energy is expected to diminish $\propto N^{-1}$ (Problem~2).  We conjecture
that, asymptotically for large $N$, the spinon interaction energy in the
2-spinon scattering states, as defined by the expression
\begin{equation}\label{eq:de2sp}
\Delta E_{2sp}^{(N)}(q)\equiv E_{2sp}^{(N)}(q) - \epsilon_{sp}(p_1) -\epsilon_{sp}(p_2)
\end{equation}
with $q=p_1+p_2$ has the form 
\begin{equation}\label{eq:esp12}
\Delta E_{2sp}^{(N)}(q) = e_{sp}(p_1,p_2)/N.
\end{equation}
The quantity $e_{sp}(p_1,p_2)$ can then be attributed to a single 2-body
collision between spinon quasi-particles traveling with momenta $p_1,p_2$.  

\begin{figure}[t!]
  \vspace*{-0.4cm}

\centerline{\epsfig{file=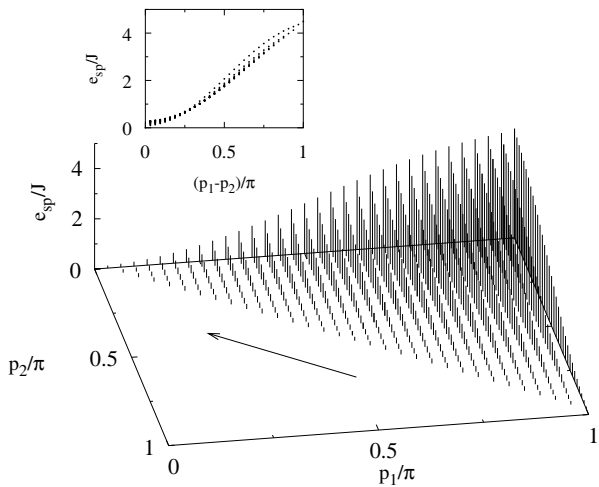,width=7.0cm,angle=-90}}
\caption{Scaled spinon interaction energy $e_{sp}(p_1,p_2)$ in the 2-spinon
  states for $N=64$ versus spinon momenta $p_1,p_2$ (3D plot) and versus
  relative spinon momentum $|p_1-p_2|$ (inset).}
\label{fig:A}
\end{figure}

The $N=64$ data for $e_{sp}(p_1,p_2)$ shown in Fig.~\ref{fig:A} indicate that
the interaction energy depends smoothly on the spinon momenta.  It is observed
to be smallest when both spinons have equal momenta. These states are near the
upper boundary of the 2-spinon continuum. For fixed momentum $p_1>p_2$, the
interaction energy is a monotonically decreasing function of $p_2$. Hence the
largest interaction energy of states at fixed wave number $q$ is realized when
the relative spinon momentum $|p_1-p_2|$ is a maximum. These states are near the
lower boundary of the 2-spinon continuum.

At fixed $|p_1-p_2|$, the spinon interaction energy depends only weakly on the
wave number $q$ of the 2-spinon state. Glancing at the data in Fig.~\ref{fig:A}
along the lines $p_1+p_2=$const. (direction of arrow), makes them
collapse into a narrow band in the plane spanned by $e_{sp}$ and $|p_1-p_2|$ as
shown in the inset to Fig.~\ref{fig:A}.

In Fig.~3 of part I we had illustrated the magnon interaction energy for
2-magnon scattering states. There the smallest interaction energy was realized
when one of two magnons had infinite wavelength $(p_2=0)$. It then acted like a
slightly rotated magnon vacuum in which the other magnon could move freely.  A
spinon, by contrast, becomes the strongest scatterer to another spinon when it
has zero momentum.

The vanishing spinon interaction energy for $N\to\infty$ does not make the
calculation of transition rates for the dynamic structure factor \eqref{eq:dssf}
a simple task. It was only recently that the exact 2-spinon part of
$S_{zz}(q,\omega)$ at $h=0$ was evaluated for an infinite chain.\cite{KMB+97} The
techniques used in that calculation, which involved generating the 2-spinon
states from the spinon vacuum by means of spinon creation operators and
expressing the spin fluctuation operator \eqref{eq:sqmu} in terms of spinon
creation operators, are not readily generalizable to $h\neq 0$ (Problem~3a).

In nonzero field, a different approach is required. The ground state must be
reinterpreted as the physical vacuum for a different species of quasi-particles.
The spectral weight of $S_{zz}(q,\omega)$ will be dominated by scattering states of
few quasi-particles from the new species, and the associated transition rates
will be calculated via Bethe ansatz.

%
\section{Quasi-particles at $\lowercase{h}\neq 0$: psinons}
\label{sec:psin}
%
Increasing the magnetic field from $h=0$ to the saturation value $h=h_S=2J$,
leaves all eigenvectors of \eqref{eq:Hh} unaltered but shifts their energies at
different rates. The ground state $|G\rangle$ changes in a sequence of level
crossings, and the magnetization increases from $M_z=0$ to $M_z=N/2$ in units of
one as discussed in part II.  Every level crossing of that sequence adds two
spinons to $|G\rangle$.

At $h\neq 0$ the number of spinons moving through the chain and scattering off each
other is of O($N$). Scattering events are frequent. The spinon interaction
energy does not vanish in the limit $N\to\infty$, because the density of spinons
remains finite. Hence the spectrum of low-lying excitations cannot be
reconstructed from the spinon energy-momentum relation \eqref{eq:spdisp}.

If we start from the ground state \mbox{$|F\rangle=|\uparrow\uparrow\cdots\uparrow\rangle$} at $h=h_S$ instead and
decrease the the magnetic field gradually, we are also facing the problem of a
runaway quasi-particle population. The state $|F\rangle$ is the vacuum of magnons, a
species of quasi-particles with spin 1. Every reverse level crossing of the
sequence considered previously adds one magnon as it removes two spinons.  In
the 2-magnon scattering states, which we had studied in part I, the role of
individual magnons was just as clearly recognizable as the role of individual
spinons in the 2-spinon scattering states depicted in Fig.~\ref{fig:0}. But in
$|G\rangle$ at $0<h<h_S$ and in all low-lying excitations from that
state, the number of magnons is macroscopic $(\propto N)$. The magnon interaction
energy remains significant, which makes it impossible to infer spectral
properties from the magnon energy-momentum relation (I5).

The ground-state wave function $|G\rangle$ at $0<h<h_S$ has spin quantum numbers
$S_T=S_T^z=M_z$ ($0 \leq M_z < N/2$) and is specified by the following set of $r=N/2-M_z$
Bethe quantum numbers:\cite{YY66a}
\begin{equation}\label{eq:IG}
\{I_i\}_G = \left\{-\frac{N}{4}+\frac{M_z}{2}+\frac{1}{2},\ldots,
\frac{N}{4}-\frac{M_z}{2}-\frac{1}{2} \right\}.
\end{equation}  
Henceforth we treat $|G\rangle$, which can be interpreted as a state containing
$N/2-M_z$ magnons or as a state containing $2M_z$ spinons, as a new physical
vacuum and describe the dynamically relevant excitation spectrum from this
reference state by modifying the uniform array (\ref{eq:IG}) of Bethe quantum
numbers systematically. For this purpose we consider the class $K_r$ of
eigenstates whose Bethe quantum numbers comprise, for $0\leq r\leq N/2$ and $0\leq m\leq
N/2-r$, all configurations
\begin{equation}\label{eq:I2msp}
-\frac{r}{2} + \frac{1}{2} -m \leq I_1 < I_2 < \cdots < I_r \leq \frac{r}{2}
- \frac{1}{2} + m.
\end{equation}
We recall from part II that the $I_i$'s must be integer valued if $r$ is odd and
half-integer valued if $r$ is even.  Every class-$K_r$ eigenstate is represented
by a real solution $\{z_i\}$ of the Bethe ansatz equation (II5),
\begin{equation}
\label{eq:bae}
N\phi(z_i) = 2\pi I_i + \sum_{j\neq i}\phi\bigl [(z_i-z_j)/2\bigr ],~ i=1,\ldots,r,
\end{equation}
with $\phi(z) \equiv 2\arctan z$.  These solutions can be obtained iteratively via (II9)
for fairly large systems.  Every class-$K_r$ state at fixed integer quantum
number $m$ $(0\leq m\leq M_z)$ can be regarded as a scattering state of $m$ pairs of
spinon-like particles, which we have named {\em psinons}.

\begin{figure}[t!]
\vspace*{0.3cm}

\centerline{\epsfig{file=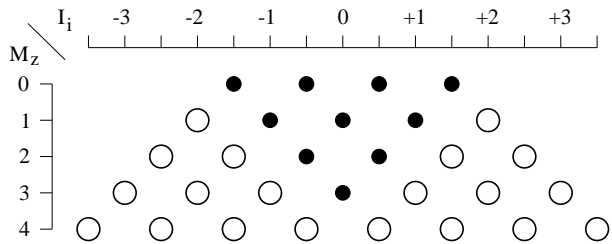,width=8.0cm}}
\caption{Psinon vacuum $|G\rangle$ for a chain of $N=8$ spins at magnetization
  $M_z=0,1,\ldots,4$. The position of the magnons $(\bullet)$ are
determined by the set \eqref{eq:IG} of $I_i$'s and the positions of the spinons
($\bigcirc$) by the vacancies across the full range of the $I_i$'s allowed by
\eqref{eq:I2msp}. }

\label{fig:1}
\end{figure}

At $M_z=0$ the psinon vacuum coincides with the spinon vacuum, both containing
$N/2$ magnons. At saturation $(M_z=N/2)$, the psinon vacuum coincides with the
magnon vacuum, a state containing $N$ spinons.  The transformation of the psinon
vacuum between the spinon vacuum and the magnon vacuum is illustrated in
Fig.~\ref{fig:1}. It displays the configuration of Bethe quantum numbers for
$|G\rangle$ in a system with $N=8$ and all values of $M_z$ realized between $h=0$ and
$h=h_S$.

The top row (spinon vacuum) corresponds to the top row in Fig.~3 of part II
(albeit for different $N$). The second row is a particular state of the 2-spinon
triplet set discussed in part II.  The third row represents the psinon vacuum at
half the saturation magnetization, $M_z/N=\frac{1}{4}$, the case we shall focus
on in this paper. Here the psinon vacuum contains twice as many spinons as it
contains magnons.  The psinon vacuum in the fourth row corresponds to the
highest-energy state of the 1-magnon branch described in part I. Trading the
last magnon for a pair of spinons saturates $M_z$ and makes the psinon vacuum
equal to the magnon vacuum (fifth row).

In the psinon vacuum $|G\rangle$, the only class-$K_r$ state with $m=0$, the magnons
form a single array flanked by two arrays of spinons. Relaxing the constraint in
(\ref{eq:I2msp}) to $m=1$ yields the 2-psinon excitations. Generically, magnons
now break into three clusters separated by the two innermost spinons, which now
assume the role of psinons. The remaining $2M_z-2$ spinons stay sidelined.

\begin{figure}[t!]
\vspace*{0.3cm}

\centerline{\epsfig{file=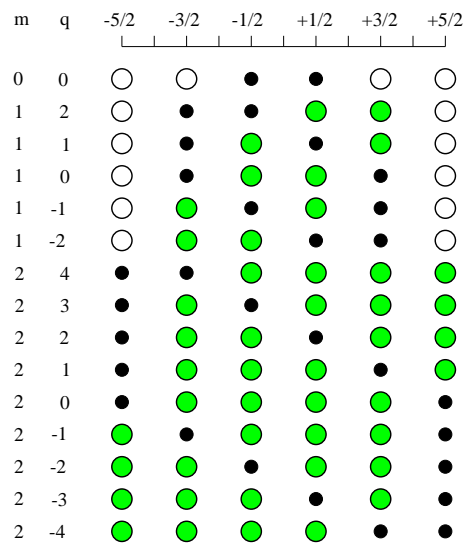,width=6.0cm}}
\caption{Psinon vacuum $(m=0)$ for $N=8,M_z=2$ and complete set of $2m$-psinon
  excitations with $m=1,2$ and wave numbers $q\equiv k-k_G$ (in units of $2\pi/N$).
  The positions of the magnons (small circles) are determined by the $I_i$'s and
  the positions of the spinons (large circles) by the $I_i$ vacancies. A subset
  of the spinons are the psinons (green circles).  }
\label{fig:2}
\end{figure}

In the 4-psinon states $(m=2)$, two additional spinons have been mobilized into
psinons. By this prescription, we can systematically generate sets of
$2m$-psinon excitations for $0\leq m\leq M_z$.  From the psinon vacuum for $N=8,
M_z=N/4=2$ depicted in the the third row of Fig.~\ref{fig:1} we can thus
generate the five 2-psinon states and the nine 4-psinon states identified
in Fig.~\ref{fig:2} 

\begin{figure}[t!]\vspace*{-0.5cm}

\includegraphics[width=7.5cm,angle=-90]{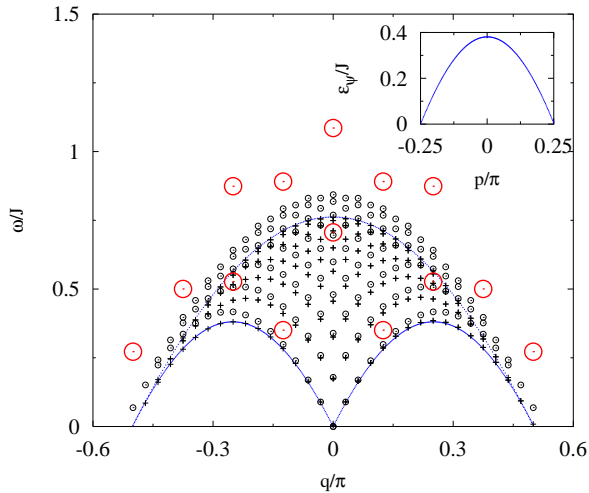}
\caption{Spectrum in $(q,\omega)$-space of all 2-psinon scattering states at $M_z/N=\frac{1}{4}$ for
  $N=16$ (large red circles) and $N=64$ (small circles), the latter in
  comparison with the corresponding (fictitious) free 2-psinon states (+). The
  spectral range of the 2-psinon continuum for $N\to\infty$ (blue lines) are inferred
  from data for $N=2048$. The inset shows the psinon energy-momentum relation.}
\label{fig:4}
\end{figure}

Figure~\ref{fig:4} shows energy versus wave number of all 2-psinon states at
$M_z/N=\frac{1}{4}$ for system sizes $N=16$ (large red circles) and $N=64$
(small circles).  In the limit $N\to\infty$, the 2-psinon states form a
continuum in $(q,\omega)$-space with boundaries shown as blue lines. The 2-psinon
spectrum is confined to $|q|\leq q_s$, where
\begin{equation}\label{eq:qs}
q_s \equiv \pi(1-2M_z/N).
\end{equation}

From the 2-psinon continuum boundaries we infer the psinon energy-momentum
relation $\epsilon_\psi(p)$ shown in the inset, by the requirement that the wave number
and the energy of every 2-psinon state for $N\to\infty$ can be accounted for by
$q=p_1+p_2$ and $\omega=\epsilon_\psi(p_1)+\epsilon_\psi(p_2)$, respectively. The two arcs that make up
the lower 2-psinon continuum boundary are then given by $\epsilon_\psi(q\pm\pi/4)$.

For finite $N$, the scattering of the two psinons in the 2-psinon states
produces an interaction energy. In Fig.~\ref{fig:4}, this energy can again be
measured by comparing the positions of the $N=64$ scattering states $(\circ)$
relative to the positions of the corresponding (fictitious) free 2-psinon states
(+). The evidence from a comparison of the $N=64$ data in Figs.~\ref{fig:1} and
\ref{fig:4} is that the psinons at $M_z/N=\frac{1}{4}$ interact more strongly
than the spinons at $M_z=0$ in chains of the same length. The interaction energy
between two psinon quasi-particles in a 2-psinon scattering state can be
investigated by the method used in Sec.~\ref{sec:2mse} for spinon
(Problem~4a). It again varies $\propto 1/N$.

\begin{figure}[t!]
\includegraphics[width=6.7cm,angle=-90]{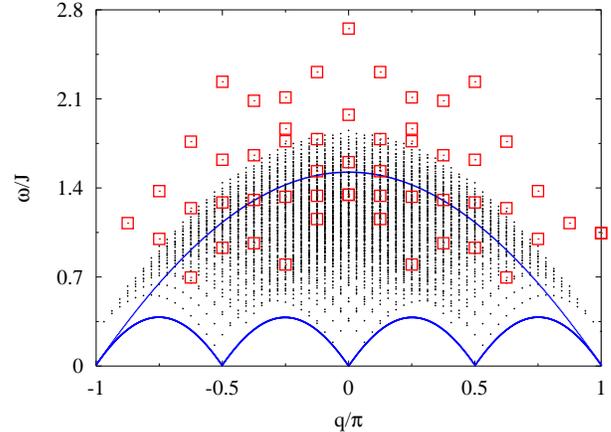}
\caption{Spectrum in $(q,\omega)$-space of all 4-psinon excitations at $M_z=N/4$ for
  $N=16$ (red squares) and $N=64$ (dots). Also shown is the range of
  the 4-psinon states for $N\to\infty$ (blue lines) inferred from $N=2048$ data.}
\label{fig:5}
\end{figure}

Mobilizing two additional spinons into psinons makes the 4-psinon scattering
states much more numerous than the 2-psinon states.  Their number grows $\propto N^4$.
In Fig.~\ref{fig:5} we have plotted all 4-psinon states for $N=16$ and $N=64$ as
well as the 4-spinon spectral range. The 4-psinon spectral threshold has the
same shape as the 2-psinon lower boundary but extended periodically over the
entire Brillouin zone. The upper 4-psinon boundary is obtained from the upper
2-psinon boundary by a scale transformation $(q\to 2q,\omega\to 2\omega)$. 

In both sets of states, the energy correction due to the psinon interaction is a
$1/N$ effect. It fades away as the scattering events become less and less
frequent in a system of increasing size. A comparison of finite-$N$ data in
Figs.~\ref{fig:4} and \ref{fig:5} shows that the finite-size energy correction
caused by the psinon interaction is stronger in the 4-psinon states
than in the 2-psinon states. The reason is the different rates at which
scattering events between psinons occur.

At $M_z=0$ the 2-spinon excitations were found to dominate the spectral weight
in the dynamic spin structure factor $S_{zz}(q,\omega)$.\cite{KMB+97} Our task here
is to determine how the spectral weight of $S_{zz}(q,\omega)$ at $M_z\neq 0$ is
distributed among the $2m$-psinon excitations. To accomplish it, we must
calculate matrix elements.

%
\section{Matrix elements}
\label{sec:me}
%
The Bethe ansatz has rarely been used for the purpose of calculating matrix
elements. Most attempts at taking this approach have been deterred by the need
of evaluating the sum ${\cal P}\in S_r$ over the $r!$ magnon permutations in the
coefficients (I28) of the Bethe eigenvectors (I27):
\begin{equation}\label{eq:psir}
 |\psi\rangle = \sum_{1\leq n_1<\ldots<n_r\leq N} a(n_1,\ldots,n_r)
 |n_1,\ldots,n_r\rangle,
\end{equation}
\vspace*{-0.5cm}
\begin{equation}
 a(n_1,\ldots,n_r) =
 \sum_{{\cal P}\in S_r}
 \exp\left( i\sum_{j=1}^r k_{{\cal P} j}n_j
 + \frac{i}{2}\sum_{i<j} \theta_{{\cal P}i{\cal P}j} \right). \nonumber
\end{equation}
The magnon momenta $\{k_i\}$ and the phase angles $\{\theta_{ij}\}$ are
related via (II4) and (II5) to the solutions $\{z_i\}$ of the Bethe ansatz
equations (\ref{eq:bae}).

In the calculation of a single matrix element, a sum ${\cal P}\in S_r$ is
evaluated many times, once for every coefficient $a(n_1,\ldots,n_r)$ of the two
eigenvectors involved. Under these circumstances, it is imperative that the
algorithm has rapid access to a table of permutations.

Table \ref{tab:T2} describes a powerful recursive algorithm that generates all
permutations of the numbers $1,2,\ldots,r$.\cite{Sedg92} These permutations are
stored in the array {\tt ip} of dimensionality $r!\times r$. The storage of this
array in the RAM requires 645kB for $r=8$, 6.53MB for $r=9$, and 72.6MB for
$r=10$.
\begin{table}[b!]
\caption{C++ code of a recursive algorithm  that
generates the $r!$ permutations of the numbers $1,2,\ldots,r$ and writes them
into an array of dimensionality $r!\times r$. Each number is stored as a
{\tt char} or {\tt short}, which is of size 2Bytes on an Intel
processor (Ref.~\cite{Sedg92}).} \label{tab:T2}    
\small
\setlength{\unitlength}{1mm}
\line(1,0){86}\\[-3mm]
\line(1,0){86}
\begin{verbatim}
void main()
{
  perm(0,r);
}
void perm(int k, int r)
{
 static long id=-1;
 static long rf=1;
 ip[0][k]=++id;
 if(id==r){
  for(int l=0;l<r;l++) ip[rf][l]=ip[0][l+1];
  rf++;
 }
 for(int t=1;t<=r;t++) if(ip[0][t]==0) perm(t,r);
 id--; 
 ip[0][k]=0;
}
\end{verbatim}
\line(1,0){86}\\[-3mm]
\line(1,0){86}
\end{table}

The computational effort of calculating any Bethe ansatz eigenvector
(\ref{eq:psir}) can be reduced considerably if we use the translational
symmetry, ${\bf T}|\psi\rangle = e^{ik}|\psi\rangle$ (see part I). It is guaranteed by the
relation
\begin{equation}\label{eq:a-translation}
a(n_1+l,\ldots,n_r+l) = e^{ikl}   a(n_1,\ldots,n_r)
\end{equation}
between sets of coefficients pertaining to basis vectors that transform into
each other under translation. Here the integers $n_i+l$ have to be used mod($N$).
Translationally invariant basis vectors have the form
\begin{equation}\label{eq:tibv}
|j;k\rangle \equiv \frac{1}{\sqrt{d_j}}\sum_{l=0}^{d_j-1}e^{ilk}|j\rangle_l,
\end{equation}
where $|j\rangle_l \equiv {\bf T}^l|j\rangle_0 =
|n_1^{(j)}-l,\ldots,n_r^{(j)}-l\rangle$  and $1\leq N/d_j\leq
N$ is an integer. The wave numbers $k$ realized in the set (\ref{eq:tibv}) are
multiples mod($2\pi$) of $2\pi/d_j$.

The set of basis vectors $|j\rangle_0 = |n_1^{(j)},\ldots,n_r^{(j)}\rangle, j=1,\ldots,d$, are the
generators of the translationally invariant basis. The set of distinct vectors
$|j;k\rangle$ for fixed $k$ is labeled $j\in{\cal J}_k \subseteq \{1,\ldots,d\}$. The rotationally
invariant subspace for fixed $N/2-r$, which has dimensionality $D =
N!/[r!(N-r)!]$, splits into $N$ translationally invariant subspaces of
dimensionality $D_k$, one for each wave number $k=2\pi n/N, n=0,\ldots,N-1$. We have
\begin{equation}\label{eq:djDk}
D = \sum_{j=1}^{d}d_j = \sum_{0\leq k< 2\pi} D_k,~~ D_k = \sum_{j\in{\cal J}_k}.
\end{equation}

The Bethe eigenvector (\ref{eq:psir}) expanded in this basis can thus be
written in the form
\begin{equation}\label{eq:psik}
|\psi\rangle = \sum_{j\in{\cal J}_k}a_j\sum_{l=0}^{d_j-1}e^{ilk}|j\rangle_l
= \sum_{j=1}^da_j\sum_{l=0}^{d_j-1}e^{ilk}|j\rangle_l, 
\end{equation}
where the $a_j \equiv a(n_1^{(j)},\ldots,n_r^{(j)})$, the Bethe coefficients of the
generator basis vectors $|j\rangle_0$, are the only ones that must be evaluated. The
last expression of (\ref{eq:psik}) holds because the Bethe coefficients $a_j$ of
all generators $|j\rangle_0$ which do not occur in the set ${\cal J}_k$ are zero
(Problem~5).

We calculate transition rates for the dynamic structure factor (\ref{eq:dssf})
in the form
\begin{equation}\label{eq:mes}
|\langle G|S_q^\mu|\lambda\rangle|^2 = \frac{|\langle\psi_0|S_q^\mu|\psi_\lambda\rangle|^2} {||\psi_0||^2 ||\psi_\lambda||^2},
\end{equation}
where $|\psi_0\rangle, |\psi_\lambda\rangle$ are the (non-normalized) Bethe eigenvectors of the ground
state and of one of the excited states from classes (i)-(vi), respectively.
The norms are evaluated as follows:
\begin{equation}\label{eq:ba-norm}
|| \psi ||^{2} = \sum_{j=1}^d d_j |a_j|^{2}.
\end{equation}
The matrix element $\langle\psi_0|S_q^\mu|\psi_\lambda\rangle$ is nonzero only if
$q=k_{\lambda}-k_{0}+2\pi\Bbb{Z}$. For the fluctuation operator $S_q^z$ it can be
evaluated in the form (Problem~6):
\begin{eqnarray}\label{eq:matrix-sq}
  && \langle \psi_0 |S_q^{z}|\psi_\lambda \rangle \!=\! 
  \frac{1}{\sqrt{N}}\sum_{j=1}^{d}\bar{a}_{j}^{(0)}a_{j}^{(\lambda)}\sum_{n=1}^{N} 
  e^{iqn}\sum_{l=0}^{d_j-1} e^{ilq} {_l}\langle j | S_n^{z} | j\rangle_{l}.
    \nonumber \\ 
\end{eqnarray}
The non-vanishing matrix elements $\langle\psi_0|S_q^\pm|\psi_\lambda\rangle$ needed for $S_{xx}(q,\omega)$
must also satisfy $q=k_{\lambda}-k_{0}+2\pi\Bbb{Z}$ and can be
reduced to somewhat more complicated expressions involving elements $_{l_0}\langle
j_0|S_n^\pm|j_\lambda\rangle_{l_\lambda}$ between basis vectors from different $S_T^z$ subspaces.

What are the memory requirements for the calculation of one such matrix element?
Three arrays are needed for the basis vectors: {\tt long b[d]}, {\tt short nd[d]}
and, {\tt ndr[d][r]}. The array {\tt b[d]} holds the bit pattern of
$|j\rangle_{0}$, {\tt nd[d]} holds the numbers $d_{j}$, and in {\tt ndr[d][r]}
we store the $r$ numbers $n_{1}^{j},\ldots,n_{r}^{j}$ belonging to
$|j\rangle_{0}$, i.e. {\tt b[j]}. The arrays {\tt nd, ndr} are not really
necessary, since we can always calculate those numbers from {\tt b[j]}, but they
help reduce the CPU-time.  Finally, the two arrays {\tt complex psi[d]} hold the
coefficients $a_{j}$.  Again, storing these numbers for multiple use cuts down
on CPU time. In Table \ref{tab:mem} we have summarized the memory requirements
for several applications to the Heisenberg model (\ref{eq:Hh}).
\begin{table}[t!]
\caption{Memory (in kB) required for the calculation of matrix elements 
           via Bethe ansatz on an Intel processor.}
\begin{center}
  \begin{tabular}{rrr|rrr|r}\hline\hline
  $N$ &$r$& $d$&{\tt ip}&{\tt b},{\tt nd},{\tt ndr}&{\tt psi0,1}&Total\\ \hline
   12 &  6&    80&    9 &                         1 &          3 &     13 \\
   14 &  7&   246&   71 &                         5 &          8 &     83 \\
   16 &  8&   810&  645 &                        18 &         26 &    689 \\
   18 &  9&  2704& 6532 &                        65 &         87 &   6684 \\
   20 & 10&  9252&72576 &                       241 &        297 &  73112 \\ \hline
   16 &  4&   116&    0 &                         2 &          3 &      6 \\
   20 &  5&   776&    1 &                        12 &         25 &     38 \\
   24 &  6&  5620&    9 &                       101 &        180 &    290 \\
   28 &  7& 42288&   71 &                       846 &       1353 &   2270 \\ 
   32 &  8& 328756&   645 &                    7232 &      10520 &  18398 \\ 
\hline \hline
  \end{tabular}
\end{center}
  \label{tab:mem}
\end{table}
With the exception of the case of $N=20,r=10$, we can compute the transition
rates on a simple Pentium-PC with at least 32MB of memory. It is also worth
mentioning that for the case $N=16,r=8$ no more than 1MB of memory is needed.

%
\section{Psinons and antipsinons}
\label{sec:psianpsi}
%
Which $2m$-psinon excitations have the largest spectral weight in $S_{zz}(q,\omega)$?
We begin with a chain of $N=16$ spins at magnetization $M_z/N=\frac{1}{4}$, and
explore the transition rates between the ground state $|G\rangle$ with $\{I_i\}_G =
\frac{1}{2}\{-3,-1,+1,+3\}$, and all $2m$-psinon excitations for $m=0,1,2,3,4$.

\begin{figure}[t1]

\centerline{\epsfig{file=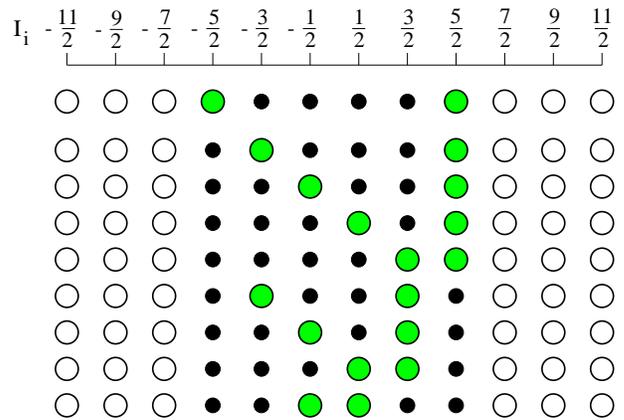,width=8.0cm}}
\caption{Psinon vacuum $|G\rangle$ for $N=16, M_z=4$ and 2-psinon states
  with $q\geq0$. The $I_i$ values are marked by the positions of the magnons (small
  circles). The spinons (large circles) mark $I_i$-vacancies. A subset of the
  spinons are called psinons (green circles).  }
\label{fig:3}
\end{figure}

First we calculate $\langle G|S_q^z|G\rangle$, which probes, for $q=0$, the ground-state
magnetization induced by the magnetic field (Problem~7). Next we
investigate the 2-psinon states. The $I_i$ configurations are shown in
Fig.~\ref{fig:3}. The first row represents the psinon vacuum with its four
magnons sandwiched by two sets of four spinons. The two innermost spinons
(marked green) become psinons when at least one of them is moved to another
position.  In the rows underneath, the psinons are moved systematically across
the array of magnons while the remaining spinons stay frozen in place. These
eight configurations describe all 2-psinon states with wave number $0\leq q\leq\pi/2$.

The $I_i$'s of the states shown in Fig.~\ref{fig:3} and their wave numbers as
inferred from (II8b) are listed in Table~\ref{tab:IV}. Solving the Bethe ansatz
equations \eqref{eq:bae} yields the ingredients needed to evaluate the energies
via (II8a) and the transition rates via \eqref{eq:matrix-sq}, which are also
listed in Table~\ref{tab:IV}.  We observe that almost all of the spectral weight
is concentrated in the lowest excitation for any given wave number (lowest red
circles in Fig.~\ref{fig:4}). In a macroscopic system, these states form the
lower boundary of the 2-psinon continuum.

\begin{table}[t!]
\caption{Ground state and 2-psinon excitations for $N=16$, $M_z=4$, and wave
  numbers $q\equiv k-k_G\geq 0$ (in units of $2\pi/N$). The ground state has
  $k_G=0$ and $E_G=-11.5121346862$.}
\begin{center}
  \begin{tabular}{cccc}\hline\hline
$2I_i$ & $k-k_G$ & $E-E_G$ & $|\langle G|S_q^z|\lambda\rangle|^2$ \\ \hline
$-3-1+1+3$ & 0 & 0.0000000000 & 1.0000000000 \\
$-5-1+1+3$ & 1 & 0.3504534152 & 0.0484825989 \\
$-5-3+1+3$ & 2 & 0.5271937189 & 0.0587154211 \\
$-5-3-1+3$ & 3 & 0.5002699273 & 0.0773592284 \\
$-5-3-1+1$ & 4 & 0.2722787522 & 0.1257902349 \\
$-5-1+1+5$ & 0 & 0.7060324808 & 0.0000000000 \\
$-5-3+1+5$ & 1 & 0.8908215652 & 0.0000064288 \\
$-5-3-1+5$ & 2 & 0.8738923064 & 0.0000312622 \\
$-5-3+3+5$ & 0 & 1.0855897189 & 0.0000000000 \\
\hline\hline
  \end{tabular}
\end{center}
\label{tab:IV}
\end{table}

When we calculate the 4-spinon transition rates $|\langle G|S_q^z|\lambda\rangle|^2$, we find that
most of their spectral weight is again carried by a single branch of
excitations. The dynamically dominant 4-psinon states for $N=16$ are the eight
lowest red squares in Fig.~\ref{fig:5}. For large $N$ they form a branch
adjacent to the 2-psinon spectral threshold.  An investigation of the
$2m$-psinon states for $m=3,4,\ldots$ shows that there exists one dynamically
dominant branch of $2m$-psinon excitations for $0<m<M_z$. All other $2m$-psinon
excitations have transition rates that are smaller by at least two orders of
magnitude at $q<\pi/2$ and still by more than one magnitude at $q\geq\pi/2$.

\begin{figure}[t!]

\centerline{\epsfig{file=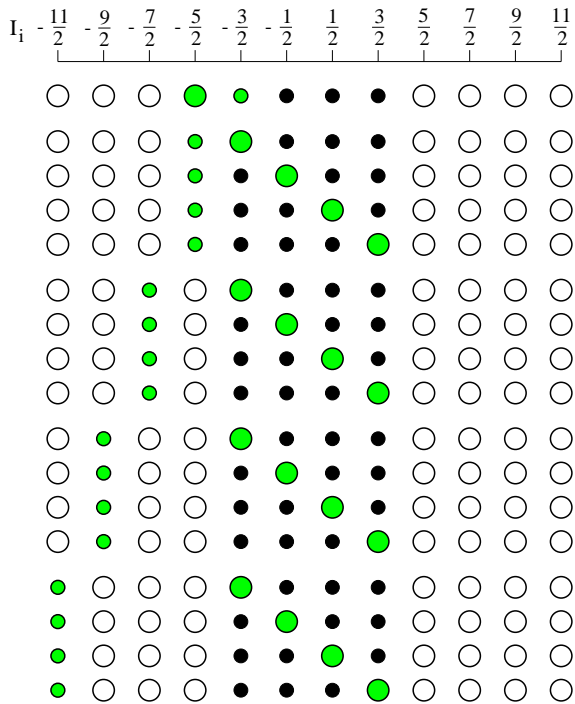,width=7.5cm}}
\caption{ Psinon vacuum $|G\rangle$ for $N=16,M_z=4$ and set of $\psi\psi^*$ states with $0\leq q< \pi$.
  The $I_i$ are given by the positions of the magnons (small circles) in each
  row. The spinons (large circles) correspond to $I_i$-vacancies. The psinon
  $(\psi)$ and the antipsinon $(\psi^*)$ are marked by a large and a small grey
  circle, respectively.}
\label{fig:6}
\end{figure}

Figure~\ref{fig:6} shows the $I_i$ configurations of the four dynamically
dominant $2m$-psinon branches for $N=16$. Each branch (at $q>0$) consists of
$N/2-M_z=4$ states. An interesting pattern emerges, which is indicative of the
nature of the relevant quasi-particles in these dynamically dominant collective
excitations. The two relevant quasi-particles are highlighted
by green circles.  

We identify one of the two quasi-particles as a psinon (large green circle) as
before and the other one as a new quasi-particle (small green circle). The
latter is represented by a hole in what was one of two spinon arrays of the
psinon vacuum. Instead of focusing on the cascade of psinons (mobile spinons)
which this hole has knocked out of the psinon vacuum, we focus on the hole
itself, which has properties commonly attributed to antiparticles. The psinon
$(\psi)$ and the antipsinon $(\psi^*)$ exist in disjunct parts of the psinon vacuum,
namely in the magnon and spinon arrays, respectively. When they cross paths at
the border of the two arrays, they undergo a mutual annihilation, represented by
the step from the second row to the top row in Fig.~\ref{fig:6}.

\begin{figure}[t!]

\centerline{\epsfig{file=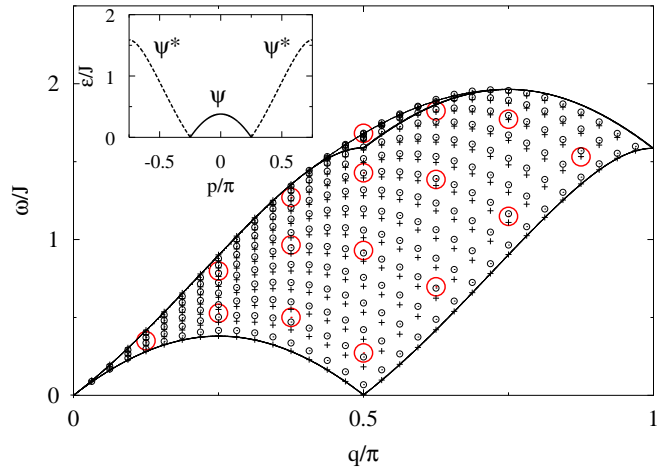,width=7.2cm,angle=-90}}
\caption{Energy versus wave number of all $\psi\psi^*$ scattering states at
  $M_{z}/N=\frac{1}{4}$ and $q\geq0$ for $N=64$ ($\circ$) in comparison with the corresponding
  free $\psi\psi^*$ states $(+)$. Also shown are the $\psi\psi^*$ states for
  $N=16$ (large red circles). The inset shows the energy-momentum relations of the
  psinon $(0\leq |p|\leq \pi/4)$ and the antipsinon $(\pi/4\leq |p|\leq3\pi/4)$.}
\label{fig:7}
\end{figure}

The large red circles in Fig.~\ref{fig:7} represent all $\psi\psi^*$ states at $q>0$
for $N=16$ as specified in Fig.~\ref{fig:6}. There are four branches ($m=1,\ldots,4$
from bottom to top) with four states each. Also shown in Fig.~\ref{fig:7} are
the $\psi\psi^*$ states for $N=64$ (small circles). The solid lines are inferred from
$\psi\psi^*$ data for $N=2048$ and represent the boundaries of the $\psi\psi^*$ continuum in
the limit $N\to\infty$.

Why have we chosen to interpret the small green circle in Fig.~\ref{fig:6} as
an antipsinon and not as a magnon? Either choice is valid but we must heed the
fact that antipsinons and magnons live in different physical vacua.

When we interpret the small green circle as a magnon, then it coexists in the
magnon vacuum with a macroscopic number of fellow magnons (small black circles).
The collective excitations must then be viewed as containing a finite density of
magnons (for $N\to\infty$), in which the magnon interaction remains energetically
significant for scattering states.  The nonvanishing interaction energy obscures
the role of individual magnons.

On the other hand, when we interpret the small green circle as an antipsinon,
then it lives in the psinon vacuum, i.e. almost in isolation. The only other
particle present is a psinon (large green circle). In the limit $N\to\infty$, the
interaction energy in a psinon-antipsinon $(\psi\psi^*)$ scattering state vanishes.

The energy-momentum relations of the two quasi-particles (see inset to
Fig.~\ref{fig:7}) can be accurately inferred from $N=2048$ data for the spectral
thresholds of the $\psi\psi^*$ states. The psinon dispersion $\epsilon_\psi(p)$ is confined to
the interval at $0\leq |p|\leq \pi/4$ (solid line) and the antipsinon dispersion
$\epsilon_{\psi^*}(p)$ to $\pi/4\leq |p|\leq3\pi/4$ (dashed line).  The lower boundary of the $\psi\psi^*$
continuum is defined by collective states in which one of the two particles has
zero energy: the antipsinon for $0\leq |q|\leq \pi/2$ and the psinon for $\pi/2\leq |q|\leq\pi$.
The upper boundary consists of three distinct segments.

For $0\leq q\lesssim 0.3935$ the highest $\psi\psi^*$ state is made up of a zero-energy
psinon with momentum $p_\psi=-\pi/4$ and an antipsinon with momentum $p_{\psi^*}=\pi/4+q$.
Likewise, for $3\pi/4\leq q\leq\pi$, the states along the upper continuum boundary are
made up of a maximum-energy antipsinon (with $p_{\psi^*}=3\pi/4$) and a
psinon with $p_\psi=-3\pi/4+q$. In these intervals, the shape of the
continuum boundary is that of $\epsilon_\psi$ or $\epsilon_{\psi^*}$.

When the two delimiting curves are extended into the middle interval, $0.3935\lesssim
q\leq3\pi/4$, they join in a cusp singularity at $q=\pi/2$. Here the maximum of
$\epsilon_\psi(p_\psi)+\epsilon_{\psi^*}(p_{\psi^*})$ subject to the constraint $p_\psi+p_{\psi^*}=q$ does not
occur at the endpoint of any quasi-particle dispersion curve. Consequently, the
$\psi\psi^*$ continuum is partially folded about the upper boundary along
the middle segment as is evident in Fig.~\ref{fig:7}.

The $(+)$ symbols in Fig.~\ref{fig:7} represent $N=64$ data of free $\psi\psi^*$
superpositions generated from the $\psi$ and $\psi^*$ energy-momentum relations. The
vertical displacement of any $(\circ)$ from the associated $(+)$ thus reflects the
interaction energy between the two quasi-particles in a $\psi\psi^*$ state. A
comparison of the $N=64$ data in Figs.~\ref{fig:3} and \ref{fig:7} shows that for
the most part the $\psi\psi^*$ interaction energy is smaller than the $\psi\psi$ interaction
energy (Problem~4b).

The lower boundary of the $\psi\psi^*$ continuum touches down to zero frequency at
$q=0$ and $q = q_s \equiv \pi/4$, which is a special case of Eq.~\eqref{eq:qs}.
Between $q_s$ and $\pi$, it rises monotonically and reaches the value $E-E_G = h$
for $N\to\infty$ (Problem~8).  A direct observation of the zero-frequency
mode at $q_s$ was made in a neutron scattering experiment on
copper benzoate.\cite{DHR+97}

When we lower $M_z$, the soft mode at $q_s$ moves to the right, the number of
$2m$-psinon branches contributing to the $\psi\psi^*$ continuum shrinks but each
branch gains additional states. At $M_z=1$ we are left with one 2-psinon branch
extending over the interior of the entire Brillouin zone. This branch is equal
to the lowest branch of 2-spinon states with dispersion $\epsilon_L(q)$, Eq.
\eqref{eq:epslu}. At $M_z=0$ the $\psi\psi^*$ excitations disappear altogether.

When we increase $M_z$ toward the saturation value, the zero-frequency mode
moves to the left, and the number of $2m$-psinon branches increases but each
branch becomes shorter. At $M_z=N/2-1$, the two-parameter set collapses into a
one-parameter set consisting of one $2m$-psinon state each for $m=1,2,\ldots,N/2-1$.
In part I these states were identified as 1-magnon excitations with dispersion
$\epsilon_1(q)=J(1-\cos q)$ (Problem~9).

%
\section{ Lineshapes}\label{sec:LS}
%

What have we accomplished thus far and what remains to be done?  We have
identified the model system and the dynamical quantity which is relevant for the
interpretation of inelastic neutron scattering data (Sec.~\ref{sec:I}).  We have
configured the ground state at $h\neq 0$ as the vacuum for psinon quasi-particles
and related it to the ground state at $h=0$, the spinon vacuum
(Secs.~\ref{sec:2mse} and \ref{sec:psin}).  We have introduced a method of
calculating matrix elements for the dynamic spin structure factor $S_{zz}(q,\omega)$
via the Bethe ansatz (Sec.~\ref{sec:me}).  We have it to identify among all the
$2m$-psinon states one continuum of collective excitations which contributes
most of the spectral weight to $S_{zz}(q,\omega)$: the $\psi\psi^*$ states
(Sec.~\ref{sec:psianpsi}). Here we use the spectral information and the
transition rates, all evaluated via Bethe ansatz, to calculate the lineshapes
relevant for fixed-$q$ scans in the neutron scattering experiment.

To finish the task, we exploit key properties of transition rates and densities
of states of sets of excitations that form two-parameter continua in
$(q,\omega)$-space for $N\to\infty$. The two parameters are quantum numbers of the $\psi$ and
$\psi^*$ quasi-particles, e.g. the positions $\nu,\nu^*$ on the $I_i$-scale of
Fig.~\ref{fig:6} of the two green circles. For $N\to\infty$, the scaled quantum numbers
$\nu/N, \nu^*/N$ become piecewise smooth functions of the physical parameters $q,\omega$.
The $\psi\psi^*$ transition rates (scaled by $N$) then turn into a continuous function
$M_{zz}^{\psi\psi^*}(q,\omega)$ and the $\psi\psi^*$ density of states (scaled by $N^{-1}$) into
a continuous function $D_{zz}^{\psi\psi^*}(q,\omega)$. The $\psi\psi^*$ spectral-weight
distribution is the product $S_{zz}^{\psi\psi^*}(q,\omega)
=D_{zz}^{\psi\psi^*}(q,\omega)M_{zz}^{\psi\psi^*}(q,\omega)$.

\begin{figure}[t!]
  \centerline{\epsfig{file=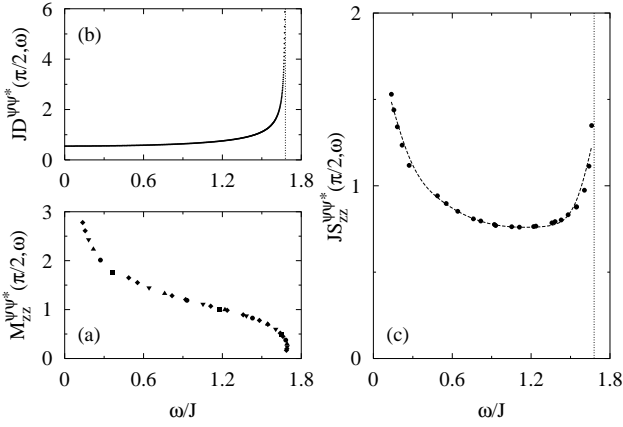,width=6.5cm,angle=-90}}

\caption{(a) Density of $\psi\psi^*$ states at $q=\pi/2$ evaluated via
  \eqref{eq:Dpsi} from Bethe ansatz data for $N=2048$. (b) Transition rates
  \eqref{eq:Mpsi} between the psinon vacuum and the $\psi\psi^*$ states at $q=\pi/2$ for
  $N=12, 16, 20, 24, 28, 32$. (c) Lineshape at $q=\pi/2$ of the $\psi\psi^*$
  contribution to $S_{zz}(q,\omega)$. All results pertain to $M_z/N=\frac{1}{4}$.}
\label{fig:8}
\end{figure}

In the following, we consider the case $q=\pi/2$, where the
$\psi\psi^*$ continuum is gapless.  The density of $\psi\psi^*$ states is generated
from $N=2048$ data of
\begin{equation}\label{eq:Dpsi}
D_{zz}^{\psi\psi^*}(q,\omega_{\nu^*}) \equiv \frac{2\pi/N}{\omega_{\nu^*+1}-\omega_{\nu^*}},
\end{equation}
where $\nu^*=m$ marks the antipsinon quantum number in the $\psi\psi^*$ continuum. The
psinon quantum number $\nu$ is adjusted to keep the wave number $q$ of the $\psi\psi^*$
state fixed.  The resulting ordered sequence of levels substituted into
\eqref{eq:Dpsi} yields the graph shown in Fig.~\ref{fig:8}(a)
(Problem~3b).

The function $D_{zz}^{\psi\psi^*}(\pi/2,\omega_{\nu^*})$ rises from a nonzero value at $\omega=0$
slowly up to near the upper boundary, where it bends into a divergence. A
log-log plot of the data near the upper band edge reveals that it is a
square-root divergence.  Finite-$N$ data over a range of system sizes for the
scaled transition rates
\begin{equation}\label{eq:Mpsi}
M_{zz}^{\psi\psi^*}(q,\omega_{\nu^*}) \equiv N|\langle G|S_{q}^{z}|\nu^*\rangle|^2
\end{equation}
with $q=\pi/2$ are shown in Fig.~\ref{fig:8}(b). They confirm the smoothness of
$M_{zz}^{\psi\psi^*}(\pi/2,\omega)$. It has a monotonic $\omega$-dependence with a divergence at
$\omega=0$ and a cusp singularity at the upper boundary $\omega_U\simeq 1.679J$ of the $\psi\psi^*$
continuum.

The product of the transition rate function and the (interpolated) density of
states is shown in Fig.~\ref{fig:8}(c). The curve fitted through the data points
represents the $\psi\psi^*$ lineshape of $S_{zz}(q,\omega)$ at $q=\pi/2$. Its most
distinctive feature is the two-peak structure caused by divergent transition
rates and a divergent density of states at the lower and upper band edges,
respectively. 
The divergence at $\omega=0$ is a power law,
$~\omega^{-\alpha}$, with an exponent that is exactly known from field theoretic
studies:\cite{Hald80} $\alpha=0.4688\ldots$.  The strength of the divergence at $\omega_U$
depends on whether or not the cusp singularity of
$M_{zz}^{\psi\psi^*}(\pi/2,\omega)$ starts from zero at $\omega=\omega_U$ (Problem~10).

The observability of this lineshape in a neutron scattering experiment also
depends on the relative $\psi\psi^*$ contribution to the integrated intensity of
$S_{zz}(\pi/2,\omega)$. With the tools developed here, we can show that the $\psi\psi^*$
states contribute at least 93\% of the spectral weight at this particular wave
number (Problem~11). A more complete analysis of the lineshapes and the
integrated intensity of $S_{zz}(q,\omega)$ for the model system at
$M_z/N=\frac{1}{4}$ can be found elsewhere.\cite{KM00}

Where do we go from here with this series of tutorial papers on the Bethe
ansatz? In the next installment, we plan on making the exchange interaction in
\eqref{eq:Hh} uniaxially anisotropic. This well set the stage for the
investigation of the following topics.  Like the magnetic field, the anisotropy
is a useful continuous parameter which brings about interesting effects in the
excitation spectrum of the spin chain. Unlike the former, the latter does affect
the wave functions of the system. The Bethe ansatz offers us a close-up look
into these changes.  Foremost among them is the transformation of 2-spinon
scattering states into 2-spinon bound states and the consequent changes of their
roles in dynamic structure factors.

\vspace*{1cm}
 
%
\section{ Problems for further study}
\label{sec:Problems}
%

(1) (a) Show that if you subject $\langle S_l^\mu(t)S_{l'}^\mu(0)\rangle$, $\mu=x,y,z$, at $T=0$
for a cyclic chain of $N$ sites to a space-time Fourier transform, $S_{\mu\mu}(q,\omega)
= N^{-1}\sum_{ll'}e^{iq(l-l')} \int_{-\infty}^{+\infty} dt\,e^{i\omega t}\langle S_l^\mu(t)S_{l'}^\mu(0)\rangle$, you
end up with expression (\ref{eq:dssf}). (b) Show that the static structure
factor $S_{\mu\mu}(q) \equiv \langle G|S_q^\mu S_{-q}^\mu|G\rangle$ as obtained from a single diagonal
matrix element is equal to the integrated intensity of (\ref{eq:dssf}),
$\int_{-\infty}^{+\infty}(d\omega/2\pi)S_{\mu\mu}(q,\omega) = \sum_m|\langle G|S_q^\mu|m\rangle|^2$ as obtained from a sum of
off-diagonal matrix elements.
  
(2) Consider the lowest-lying 2-spinon excitation at $q=\pi/2$.  Find the pattern
of its Bethe quantum numbers $I_i$ for arbitrary $N$. Use Eq.~(II9) to calculate
the excitation energy $E_{\pi/2}(N)$ of that state for a range of system sizes.
Plot the spinon interaction energy $E_{\pi/2}(N)-\pi J/2$ versus $1/N$ to verify the
$N$-dependence suggested in the text. Repeat the same task for the lowest
2-spinon excitation at $q=\pi$. Here the reference energy is zero. Verify the
exact result\cite{HHM95} $E_\pi(N)\sim\alpha/N$, $\alpha=\pi^2/2$ for this excitation via
extrapolation.
  
(3) (a) Use the result (II30), $\epsilon(q,\bar{q}) =\pi J|\sin\frac{1}{2}q\cos\bar{q}|$,
$0\leq\bar{q}\leq q$, for the 2-spinon spectrum at some wave number $0\leq q\leq\pi$ to
calculate the 2-spinon density of states analytically via $D(q,\omega) =
\int_0^qd\bar{q}\delta\left(\omega-\epsilon(q,\bar{q})\right)$. The result $D(q,\omega) =
\left[\epsilon_U^2(q)-\omega^2\right]^{-1/2}$ for $\epsilon_L(q)<\omega\epsilon_U(q)$ is then to be multiplied
by the exact transition rate function $M(q,\omega)$ derived in
Ref.~\onlinecite{KMB+97} to produce the exact 2-spinon part of $S_{zz}(q\omega)$. (b)
Reproduce the analytic result for $D(\pi,\omega)$ computationally from finite-$N$ data
via Eq.~\eqref{eq:Dpsi} with $\nu^*$ now labelling the 2-spinon states at $q=\pi$ in
order of increasing energy.
  
(4) (a) For sufficiently large $N$ the psinon interaction energy in the 2-psinon
scattering states is of the form $\Delta E_{2\psi}^{(N)}(q) \simeq e_{\psi}(p_1,p_2)/N$, where
$q=p_1+p_2$ and $e_{\psi}(p_1,p_2)$ depends smoothly on the psinon momenta
$p_1,p_2$. Since we do not know the psinon energy-momentum relation $\epsilon_\psi(p)$
analytically, use the 2-psinon lower boundary for $N^*\gg N$ as an approximation
for $\epsilon_\psi(p\pm\pi/4)$ in $\Delta E_{2\psi}^{(N)}(q)\equiv E_{2\psi}^{(N)}(q) -\epsilon_\psi(p_1) -\epsilon_\psi(p_2)$.
Plot $N\Delta E_{2\psi}^{(N)}(q)$ versus $(p_1,p_2)$ for 2-psinon states of systems with
$N^*\gg N \gg 1$ judiciously chosen. Compare the properties of the newly found
function $e_{\psi}(p_1,p_2)$ with those of the function $e_{sp}(p_1,p_2)$
established in Fig.~\ref{fig:A} for spinons. (b) Carry out the same procedure
for the $\psi\psi^*$ states. The lowest branch of states is the same as in (a).
Compare the trends in quasi-particle interaction energies with increasing energy
of the 2-psinon states and $\psi\psi^*$ states.

(5) Identify a complete set of generators $|j\rangle_0, j=1,\ldots,d$ for the
(\ref{eq:tibv}) in the subspace $N=6, r=3$. Verify that the $d_j$ associated
with these generators add up to $D$. Establish the sets ${\mathcal J}_k$ for
$k=0, \pi/3, \ldots, 5\pi/3$. Show that the $D_k$ add up to $D$. Calculate the Bethe
coefficients $a_j$ of the $d=4$ generators for all $D=20$ solutions of the Bethe
ansatz equations by using the data from Table IV of part I. Show that all $a_j$
with $j\notin{\mathcal J}_k$ vanish.
  
(6) Use the results of Problem~5 to calculate (for $N=6$, $r=3$) the structure
factor $\langle G|S_q^\mu S_{-q}^\mu|G\rangle$ and the transition rates $|\langle G|S_q^\mu|m\rangle|^2$. Show
that these data satisfy the general result of Problem~1(b).

(7) Show that $|\langle G|S_{q=0}^z|G\rangle|^2 = M_z^2/N$ and use this result to test your
computer program which calculates transition rates from Bethe wave functions.

(8) The $2m$-psinon branch $(0<m\leq M_z)$ of the $\psi\psi^*$ continuum at $q>0$ is
specified by the $r=N/2-M_z$ Bethe quantum numbers $I_1 = -N/4 +M_z/2
+\frac{1}{2} -m$, $-N/4 +M_z/2 +\frac{1}{2} \leq I_2 <\ldots < I_R \leq N/4 -M_z/2
-\frac{1}{2}$. (a) Identify the Bethe quantum numbers of the $\psi\psi^*$ state with
the lowest energy above the ground state $|G\rangle$.  Show that the wave number of
this state is $k-k_G = \pi(1-2M_z/N)$. Show by numerical extrapolation of
finite-$N$ data that its excitation energy tends to zero as $N\to\infty$. (b) Identify
the Bethe quantum numbers of the sole $\psi\psi^*$ state with $q=\pi-2\pi/N$.  Plotting
the excitation energy of this state versus $M_z/N$ yields a set of data points
that converges, as $N\to\infty$, toward the inverse magnetization curve $h(M_z)$.
  
(9) Use the exact 1-magnon wave functions (I6) with Bethe coefficients (I8) in
expressions (\ref{eq:mes})-(\ref{eq:matrix-sq}) to calculate the $\psi\psi^*$
transition rates $|\langle G|S_q^z|\lambda\rangle|^2$ for $M_z=N/2-1$ and arbitrary $N$. Note the
different $N$-dependence for $q=0$ and $q\neq 0$. Show that the relative $\psi\psi^*$
spectral weight in $S_{zz}(q,\omega)$ is 100\%, but the absolute intensity for $q\neq0$
is only of O($N^{-1}$).

(10) (a) Use the lowest-energy excitation for $N=12,16,\ldots,32$ to perform a
nonlinear fit of the expression $M_{zz}^{\psi\bar{\psi}}(\pi/2,\omega_{\bar{\nu}}) \sim a_1 +
a_2\omega^{-\alpha}$ and compare the exponent value thus obtained with the field-theoretic
prediction quoted in the text. (b) Use all data at $\hbar\omega/J>1$ to fit the
expression $M_{zz}^{\psi\bar{\psi}}(\pi/2,\omega_{\bar{\nu}}) \sim b_1 + b_2(\omega_U-\omega)^{\beta}$. Perform
an alternative fit in which $b_1$ is forcibly set equal to zero.

(11) For fixed $N$, the integrated intensity is the expectation
  value $\langle G|S_{\pi/2}^zS_{\pi/2}^z|G\rangle$ and the $\psi\psi^*$ part thereof the sum of
  transition rates $\sum_{\nu^*=1}^{N/4}|\langle G|S_{\pi/2}^z|\nu^*\rangle|^2$ as worked out in
  Problem~1. Evaluate these quantities from the Bethe wave functions
  for $N=12,16,\ldots$ and extrapolate the ratio find the relative spectral weight of
  the $\psi\psi^*$ excitations in $S_{zz}(\pi/2,\omega)$.

\vspace*{1cm}

%
\acknowledgments
%
Financial support from the URI Research Office (for G.M.) and from the DFG
Schwerpunkt \textit{Kollektive Quantenzust{\"a}nde in elektronischen 1D
  {\"U}bergangsmetallverbindungen} (for M.K.)  is gratefully acknowledged.

\vspace*{1cm}


\end{document}